\documentclass[8.5pt,twoside,twocolumn]{article}
\usepackage[super,sort&compress,comma]{natbib}
\usepackage{widetext}

\usepackage[version=3]{mhchem}
\usepackage{times,mathptmx}
\usepackage{sectsty}
\usepackage{balance}

\usepackage{graphicx} 
\usepackage{lastpage}
\usepackage[format=plain,justification=raggedright,singlelinecheck=false,font=small,labelfont=bf,labelsep=space]{caption}
\usepackage{fancyhdr}
\usepackage{amssymb,amsmath}
\begin{document}

\thispagestyle{plain}
\fancypagestyle{plain}{
\fancyhead[L]{\includegraphics[height=8pt]{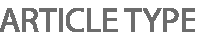}}
\fancyhead[C]{\hspace{-1cm}\includegraphics[height=20pt]{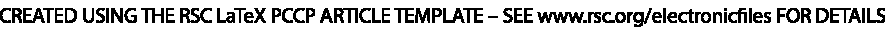}}
\fancyhead[R]{\includegraphics[height=10pt]{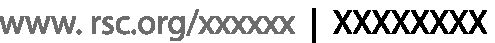}\vspace{-0.2cm}}
\renewcommand{\headrulewidth}{1pt}}
\renewcommand{\thefootnote}{\fnsymbol{footnote}}
\renewcommand\footnoterule{\vspace*{1pt}%
\hrule width 3.4in height 0.4pt \vspace*{5pt}}

\setcounter{secnumdepth}{5}

\makeatletter
\def\subsubsection{\@startsection{subsubsection}{3}{10pt}{-1.25ex plus -1ex minus -.1ex}{0ex plus 0ex}{\normalsize\bf}}
\def\paragraph{\@startsection{paragraph}{4}{10pt}{-1.25ex plus -1ex minus -.1ex}{0ex plus 0ex}{\normalsize\textit}}
\renewcommand\@biblabel[1]{#1}
\renewcommand\@makefntext[1]%
{\noindent\makebox[0pt][r]{\@thefnmark\,}#1}
\makeatother
\renewcommand{\figurename}{\small{Fig.}~}
\sectionfont{\large}
\subsectionfont{\normalsize}

\fancyfoot{}
\fancyfoot[LO,RE]{\vspace{-7pt}\includegraphics[height=9pt]{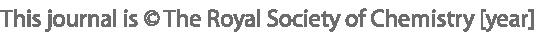}}
\fancyfoot[CO]{\vspace{-7.2pt}\hspace{12.2cm}\includegraphics{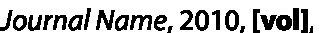}}
\fancyfoot[CE]{\vspace{-7.5pt}\hspace{-13.5cm}\includegraphics{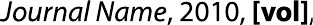}}
\fancyfoot[RO]{\footnotesize{\sffamily{1--\pageref{LastPage} ~\textbar  \hspace{2pt}\thepage}}}
\fancyfoot[LE]{\footnotesize{\sffamily{\thepage~\textbar\hspace{3.45cm} 1--\pageref{LastPage}}}}
\fancyhead{}
\renewcommand{\headrulewidth}{1pt}
\renewcommand{\footrulewidth}{1pt}
\setlength{\arrayrulewidth}{1pt}
\setlength{\columnsep}{6.5mm}
\setlength\bibsep{1pt}

\twocolumn[
 \begin{@twocolumnfalse}
\noindent\LARGE{\textbf{Effects of Precursor Topology and Synthesis under Crowding Conditions on the Structure of Single-Chain Polymer Nanoparticles}} 
\vspace{0.6cm}

\noindent\large{Maud Formanek,\textit{$^{a}$} Angel J. Moreno,$^{\ast}$\textit{$^{a}$}\textit{$^{b}$} } 
\vspace{0.5cm}

\noindent\normalsize{By means of molecular dynamics simulations we investigate the formation
of single-chain nanoparticles through intramolecular cross-linking of polymer chains, in the presence
of their precursors acting as purely steric crowders in concentrated solution. In the case of the linear precursors, the structure of the resulting SCNPs is weakly affected by the density at which the synthesis is performed. Crowding has significant effects if ring precursors are used: higher concentrations lead to the formation of SCNPs with more compact and spherical morphologies. Such SCNPs retain in the swollen state (high dilution) the crumpled globular conformations adopted by the ring precursors in the crowded solutions. 
Increasing the concentration of both the linear and ring precursors up to 30 \% leads to faster formation of the respective SCNPs.}
\vspace{0.5cm}
 \end{@twocolumnfalse}
  ]

\footnotetext{\textit{$^{a}$Centro de F\'isica de Materiales (CSIC, UPV/EHU) and Materials Physics Center MPC, Paseo Manuel de Lardizabal 5, E-20018 San Sebasti\'an, Spain.  E-mail: angeljose.moreno@ehu.eus}}
\footnotetext{\textit{$^{b}$Donostia International Physics Center (DIPC), Paseo Manuel de Lardizabal 4, E-20018 San Sebasti\'an, Spain. }}





\section{Introduction}

Single-chain nanoparticles (SCNPs) are an emergent class of soft nano-objects of molecular size of 
5-20 nm \cite{Altintas2012,lyon2015,reviewcsr,mavila2016,hanlon2016}. They are synthesized, generally at high dilution 
($\sim 1$~mg/mL), through purely
intramolecular cross-linking of the reactive functional groups of single polymer precursors.
A growing interest is being devoted in recent years to develop a SCNP-based technology 
with multiple applications in catalysis \cite{terashima2011,perez2013endowing,huerta2013,tooley2015}, nanomedicine \cite{hamilton2009,sanchez2013design},  bioimaging \cite{perezbaena2010,bai2014}, biosensing \cite{gillisen2012single}, or
rheology \cite{Mackay2003,Arbe2016,Bacova2017} among others. 
A recent review of the state-of-the-art in fundamentals and applications of SCNPs can be found in Ref.~\cite{PomposoSCNPbook}.

Though advanced methods have been recently introduced to produce compact SCNPs in good solvent \cite{chao2013,perezbaena2014thiol},
the latter are more the exception than the rule. 
Recent works by small-angle neutron and X-ray scattering (SANS and SAXS) have indeed revealed that, in general,
SCNPs in good solvent and high dilution are open sparse objects\cite{sanchez2013design,perez2013endowing,Sanchez-Sanchez2013a,Moreno2013}. Their size $R$ scales with the polymerization degree $N$ as  $R\sim N^{\nu}$, 
with an average exponent $\nu \approx 0.5$ (see the compilation of literature results in Ref.~\cite{Pomposo2014a}).
This observation is rather different from the limit
of globular spherical objects ($\nu = 1/3$). 
Computer simulations have elucidated the underlying physical mechanism for such morphologies \cite{Moreno2013}. 
Namely, the precursors are self-avoiding chains (scaling with the Flory exponent $\nu_{\rm F} = 0.59$) \cite{Rubinstein2003} in the standard good solvent conditions of synthesis.
In such conditions their open conformations promote bonding between reactive groups that are separated by short contour distances.
This mechanism is inefficient for compaction of the SCNP, irrespective of the degree of cross-linking \cite{Moreno2013}. Compaction is instead favoured by cross-linking events connecting precursor segments separated by long contour distances. Clearly, such events
are very unfrequent in self-avoiding chains, and the number of formed long-range loops is not sufficient to achieve
global compaction of the SCNP. Though the cross-linking process of a same polymer precursor produces topologically polydisperse SCNPs, the resulting distribution is dominated by sparse morphologies \cite{Moreno2013,LoVerso2014}.

The scaling behavior of SCNPs in good solvent at high dilution ($\nu \approx 0.5$) is similar to that found
for intrinsically disordered proteins (IDPs) {\it in vitro} \cite{marsh2010,hofmann2012,bernado2012,wendell2014}.
This behavior is intermediate between that of denatured unfolded ($\nu \approx \nu_{\rm F}$) and globular folded
proteins ($\nu \approx 1/3$), in analogy with the observation for SCNPs (intermediate between the limits of self-avoiding and collapsed globular chains). In general, IDPs are not fully disordered polymers that can be represented as linear chains. 
Most of them have some degree 
of secondary structure. Thus, IDPs are topologically polydisperse and exhibit 
very different degrees of disorder and compactness \cite{receveur2012,habchi2014,vanderlee2014}.
In a recent work \cite{Moreno2016JPCL}, structural analogies between SCNPs and IDPs have been explored.
Simulations have revealed that, despite lacking of ordered regions, SCNPs still show weakly deformable compact `domains' (disordered analogues of the IDP domains)
connected by flexible disordered segments \cite{Moreno2016JPCL}.
A criterion for the degree of internal disorder, based on the molecular asphericity and the size of the domains, has allowed to investigate separately the specific effect of the steric crowding on the collapse behaviour of each SCNP
in concentrated solution. Increasing the density of the solution leads to collapse from self-avoiding to Gaussian conformations only in the limit
of fully disordered SCNPs \cite{Moreno2016JPCL} or linear chains \cite{kang2015}. In general, as a consequence of their molecular topology with permanent loops, SCNPs in crowded solutions adopt crumpled globular conformations similar to those
found for melts of ring polymers \cite{Halverson2011, Reigh2013, Goossen2014}. This observation, confirmed by SANS experiments \cite{Moreno2016JPCL}, in SCNPs ---a system showing structural analogies with IDPs but free of specific interactions --- suggests
the former general scenario for the contribution of purely steric crowding to the conformations of IDPs {\it in vivo} (cell environments
with concentrations of 10-40 \% \cite{Theillet2014}). 

The work of Ref.~\cite{Moreno2016JPCL} investigated the effect of steric crowding on the conformations of SCNPs obtained by synthesis from linear precursors {\it at high dilution}. As aforementioned, the resulting SCNPs were topologically sparse objects that collapsed to crumpled globular conformations in concentrated solutions.
In this article we follow the inverse procedure. The simulations
of the SCNP formation are carried out in the presence of inert crowders, 
at concentrations well beyond (5-15 times) the overlap density. The crowders are the same precursors. Only a dilute fraction of them is functionalized and forms SCNPs, so that intermolecular cross-linking is negligible.  
We investigate the cases of linear and ring precursors.
Unlike for the synthesis at high dilution, the conformations of the precursors are not self-avoiding in the former concentrated solutions, but Gaussian or crumpled globular for the linear and ring case, respectively. After completing cross-linking we remove the crowders and characterize in the swollen state, at high dilution, the size and shape of the SCNPs obtained in the crowded solution. 
In the case of the SCNPs obtained from linear precursors, their structure is weakly affected by the density at which the synthesis is performed. However, crowding has significant effects if ring precursors are used: higher concentrations lead to the formation of SCNPs with more compact and spherical structures. The swollen SCNPs retain at high dilution the crumpled globular conformations adopted by their ring precursors in the crowded solutions. 
We have also investigated the effect of steric crowding on the kinetics of cross-linking.
Though an ultimate crossover to deceleration is expected at higher densities,
increasing the concentration of both the linear and ring precursors up to 30 \% leads to faster formation of the respective SCNPs.

The article is organized as follows. In Section 2 we present details of the model and the simulation method. In Section 3 we characterize and discuss the effect of the precursor topology and the synthesis
under crowding on the structure of the resulting SCNPs. In this section we also analyze 
and discuss the kinetics of cross-linking as a function
of the concentration. Conclusions are given in Section 4.

\section{Model and simulation details}
We simulate the precursors as well as the synthesized SCNPs via a coarse-grained bead-spring model\cite{Kremer1990} in good solvent conditions. Following the well-established Kremer-Grest model\cite{Kremer1990}, the non-bonded interactions between any two given monomers (both of the reactive and non-reactive kind) are modeled by a purely repulsive Lennard-Jones (LJ) potential, 
\begin{equation}
U^{\rm LJ}(r) = 4\epsilon \left[ \left(\frac{\sigma}{r}\right)^{12} -\left(\frac{\sigma}{r}\right)^{6} +\frac{1}{4}\right] \, ,
\end{equation}
with a cutoff distance $r_{\rm c} = 2^{1/6}\sigma$, at which both the potential and the corresponding forces are continuous. Furthermore, connected beads along the chain contour, as well as cross-linked beads after synthesis of the SCNPs, interact via a finitely extensible nonlinear elastic (FENE) potential, 
\begin{equation}
U^{\rm FENE}(r) = - \epsilon K_{\rm F} R_0^2 \ln \left[ 1 - \left(  \frac{r}{R_0}\right)^2 \right] \, , 
\label{eq:fene}
\end{equation}
with $K_{\rm F} = 15$ and $R_0 = 1.5$. This combination of LJ and FENE potentials guarantees chain uncrossability, limits the fluctuation of bonds and mimics good solvent conditions. In what follows, we employ standard LJ units, $\epsilon = \sigma = m = 1$ (with $m$ being the monomer mass), setting the energy, length and time ($\tau = \sqrt{\sigma^2m/\epsilon}$) scales, respectively.

The number of monomers in the precursor molecules is $N = 160$ for the linear chains and $N = 250$ for the ring polymers, which correspond to the same radii of gyration, $R_{\rm g}\approx 10\sigma$, of those molecules at high dilution. The fraction of functional reactive  groups, $f = N_{\rm r} / N = 0.25$ is the same for both linear and ring precursor molecules (i.e. $N_{\rm r} = 40$ for chains and $N_{\rm r} = 62$ for rings).
These reactive monomers are distributed randomly across the polymer, with the only constraint being that the placement of consecutive functional groups is forbidden, in order to prevent trivial cross-links. 

We perform Langevin dynamics simulations at a fixed temperature $T = \epsilon/k_{\rm B} = 1$. The equations of motion are integrated using the velocity-Verlet algorithm with a time step of $\Delta t = 0.01\tau$, following the impulse approach as put forward in, e.g., refs \cite{Izaguirre2001} and \cite{Smith2009}. The size of our simulation box is $V = L^3 = (100\sigma)^3$ and we run simulations with three different monomer densities $\rho = N N_{\rm c}/L^3 = \lbrace 0.1, 0.2, 0.3\rbrace$, where $N_{\rm c}$ is the total number of molecules in the system. This range of densities lies well over the overlap density (5-15 times) for both precursor molecules, $\rho^* = N/\langle D_{\rm g0}\rangle^3 \sim \mathcal{O}(10^{-2})$, where $D_{\rm g0}$ is the diameter of gyration at $\rho \rightarrow 0$. In the case of the rings, they are initially constructed
as planar objects and placed in the simulation box in positions that prevent concatenation.
In each independent run, two precursor molecules with randomly distributed reactive groups are present, which corresponds to a density of reactive molecules comparable to that used in the standard synthesis protocol at high dilution ($\rho \approx 10^{-4}-10^{-3}$). The rest of the molecules are non-reactive precursors of the same topology (i.e. rings or linear chains) and polymerization degree $N$ as the reactive precursors.  

\begin{figure}
\centering
  \includegraphics[height=4.5cm]{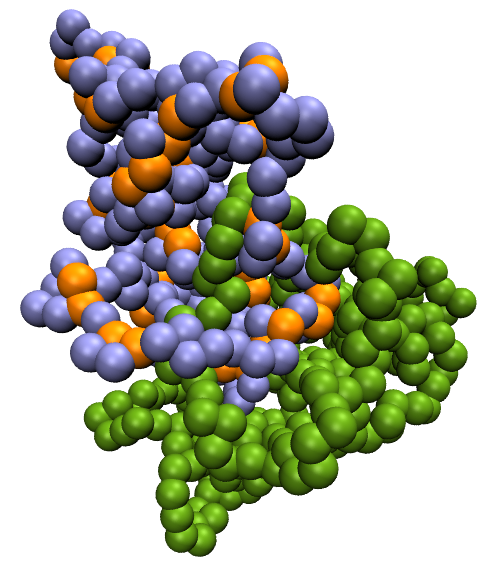}
  \caption{Example of a concatenation of a ring SCNP (blue) with a non-reactive ring polymer (green) due to the cross-links formed while the two rings are interpenetrating each other. Reactive monomers are coloured in orange.}
  \label{fgr:concatenation}
\end{figure}

Our simulation protocol consists of three steps. First the polymers are equilibrated over several million time steps at the desired density without allowing the reactive groups to cross-link. 
In the next step, SCNPs are synthesized by starting the cross-linking process in the reactive precursors. A detailed description of its implementation can be found in ref \cite{Moreno2013}. Briefly, cross-links in the synthesis runs are monofunctional and irreversible. Therefore, two reactive monomers can form a mutual bond only if they are not bonded to any other reactive groups and are separated by less than the "capture distance" $r_b = 1.3\sigma$. Whenever more than one cross-link is possible for any given monomer at any given time, one of the candidate bonds is chosen at random. Once a bond is formed, the two involved monomers interact via the FENE potential introduced in eqn \ref{eq:fene} for the remainder of the simulation. For each density of crowder molecules considered, a total of 100 independent boxes were simulated, leading to cross-linking of 200 reactive precursors through the former scheme. Intermolecular cross-linking of the reactive precursors was marginal. For each density it was only observed, at most, in 2 of the 100 independent cross-linking runs, 
which were excluded from the statistical analysis. Furthermore, despite being initially unconcatenated, for the ring polymers it is possible that the cross-linking process leads to concatenations between the SCNP and non-reactive rings (see an example in Fig.~\ref{fgr:concatenation}), since ring polymers at high densities, although adopting compact conformations, still exhibit significant interpenetration\cite{Halverson2011}. The few cases ($\leq 2\%$) in which concatenation indeed occurred where also excluded from the statistical analysis.  

After the cross-linking procedure is complete, the crowders are removed and the synthesized SCNPs are simulated in the swollen state, at infinite dilution ($\rho = 0$) in order to later compare their structure to those synthesized without the effects of crowding. To achieve efficient thermalization and limit temperature fluctuations at $\rho =0$, all the fully cross-linked SCNPs are placed in the same box -- and thus coupled to the same thermal bath, -- but are propagated independently by switching off the intermolecular interactions. After equilibration under these new conditions, simulations are further extended over several million time steps to accumulate configurations for statistical time-averages. 

\section{Results and discussion}
\begin{figure}
\centering
  \includegraphics[height=5.5cm]{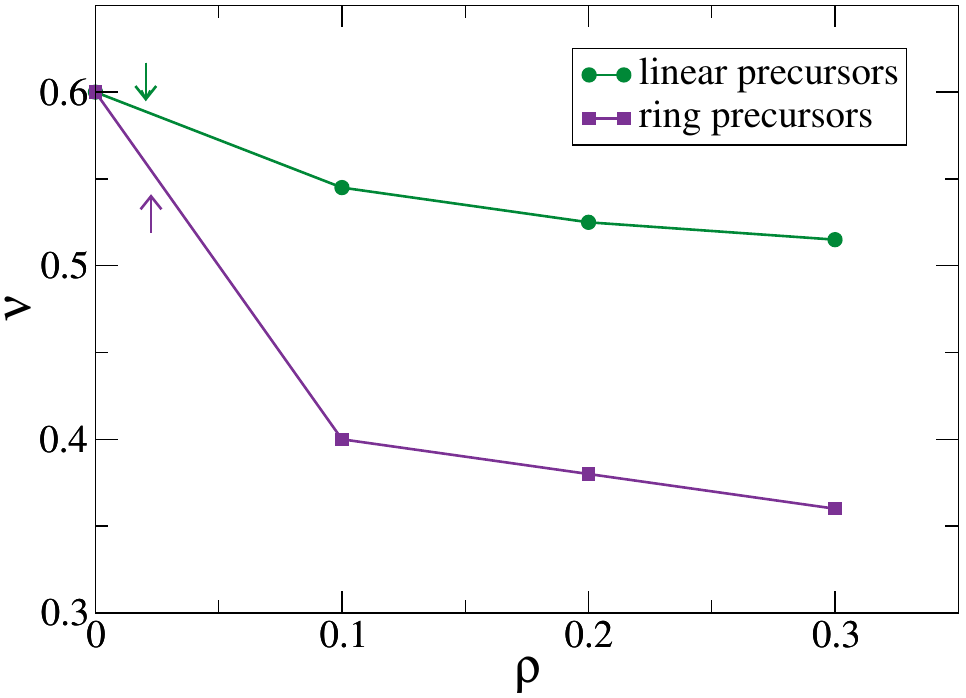}
  \caption{Scaling exponents of the two different precursor types at densities ranging from $\rho = 0$ to $\rho = 0.3$. Arrows indicate the overlap density $\rho^* = N/ \langle D_{g0}\rangle^3$.}
  \label{fgr:scaling-precursor}
\end{figure}

Fig.~\ref{fgr:scaling-precursor} shows the scaling exponents $\nu$ of the precursor molecules -- calculated from fitting the intramolecular form factors to a power law $w(q) \sim q^{-1/\nu}$ in the fractal regime (see below) -- at various densities. At infinite dilution both the linear chains and the rings show the expected Flory exponent $\nu \approx \nu_{\rm F} \approx 0.59$ for self-avoiding polymers.
At the highest density considered, $\rho = 0.3$, the linear precursors approach the scaling exponent of linear polymer melts (Gaussian chains, $\nu = 0.5$)\cite{Doi1986, Rubinstein2003}, while the ring precursors exhibit a scaling exponent  $\nu = 0.36$, which suggests highly collapsed conformations and is consistent with the scaling exponents found for ring polymers in melts, where they adopt crumpled globular conformations\cite{Halverson2011, Reigh2013, Goossen2014}. 

We investigate the effect of crowding on the resulting topology and structure of SCNPs by analyzing their size and their shape by means of the radius of gyration and the asphericity. These can be obtained from the eigenvalues $\lambda_1 \geq \lambda_2\geq \lambda_3$ of the gyration tensor,\cite{Solc1973} which is defined as: 
 
\begin{equation}
T_{\alpha\beta} = \frac{1}{N^2} \sum_{i=1}^N (r_{i\alpha} - r_{\alpha}^{\rm cm}) (r_{i\beta} - r_{\beta}^{\rm cm}) \, , 
\label{eq:gyrtens}
\end{equation}
where $r_{i\alpha}$ is the $\alpha$-th cartesian component of the position of monomer $i$ within a given polymer, and $r_{\alpha}^{\rm cm}$ is the same cartesian component of the center-of-mass of that polymer. The trace of the gyration tensor corresponds to the squared radius of gyration, i.e., 
\begin{equation}
R_g = \left(\lambda_1 + \lambda_2 + \lambda_3\right)^{\frac{1}{2}} \, , 
\label{eq:rgyr}
\end{equation}
while the asphericity is calculated as:\cite{Theodorou1985, Rudnick1986, Rudnick1987} 
\begin{equation}
a = \frac{(\lambda_2-\lambda_1)^2 + (\lambda_3-\lambda_1)^2 + (\lambda_3-\lambda_2)^2}{2(\lambda_1+\lambda_2+\lambda_3)^2} \, .
\label{eq:asph}
\end{equation}
The asphericity ranges from $0$ for objects with spherical symmetry to $1$ for a 1-dimensional object ($\lambda_2 = \lambda_3 = 0$). 

\begin{figure}
\centering
  \includegraphics[height=7.5cm]{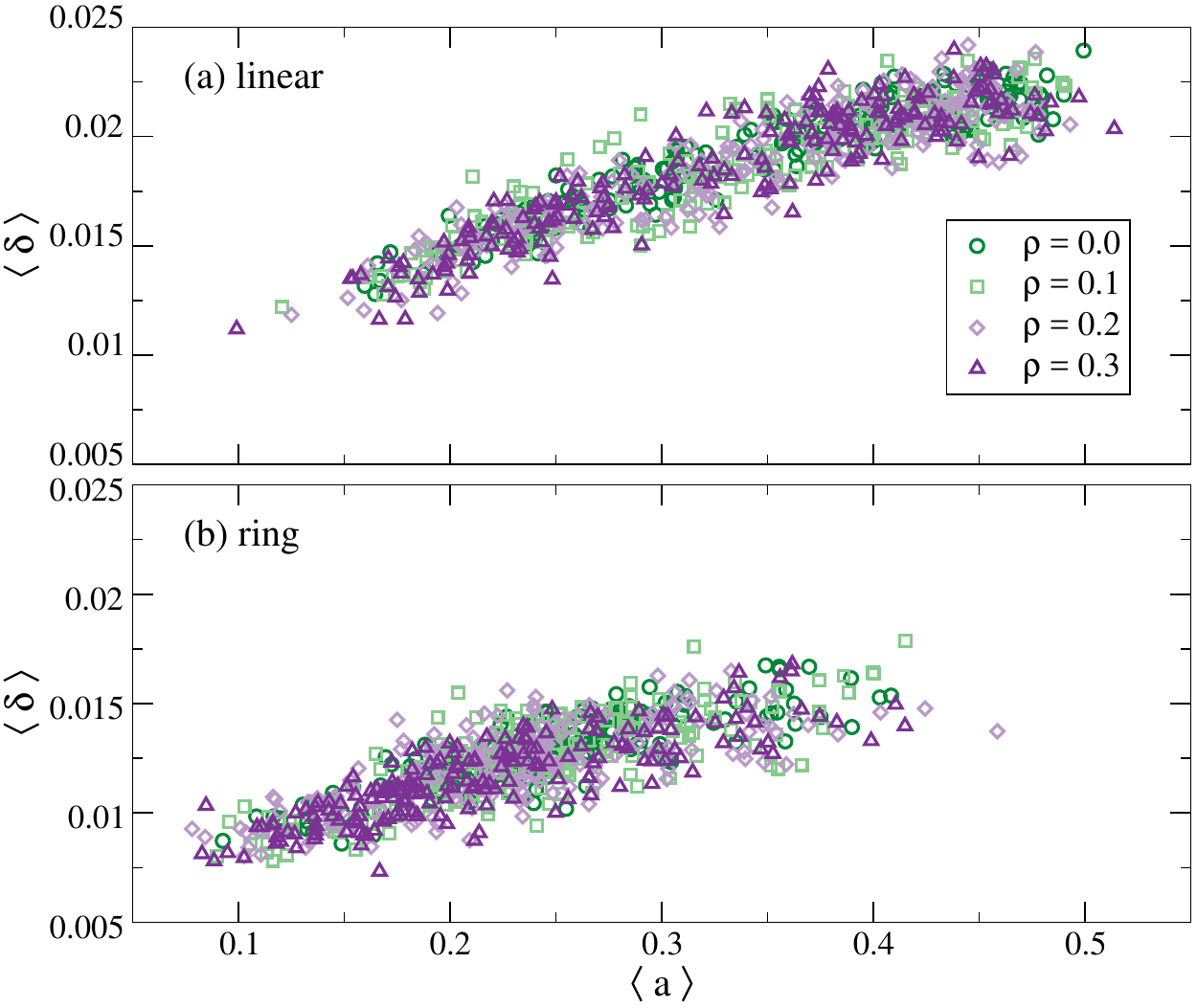}
  \caption{Relative fluctuation $\delta$, at infinite dilution, of SCNPs versus their asphericity $a$, for SCNPs synthesized from linear (a) and ring (b) precursors at various densitites. Brackets denote time-averages over the trajectory of the individual SCNP. Therefore, each point in the plot corresponds to the time-averaged value of an individual SCNP. Symbol codes have the same meaning in both panels.}
  \label{fgr:asph-fluc}
\end{figure}

Since the stochastic cross-linking process leads to a high topological and structural polydispersity among the resulting SCNPs\cite{Moreno2016JPCL}, we calculate all characteristics, such as $\langle a\rangle$, for individual SCNPs, where angular brackets denote time averages over the course of the simulation of the swollen SCNPs at $\rho = 0$. The asphericity shows a strong correlation with the internal fluctuation of the SCNP, defined as: 
\begin{equation}
\delta = \left(\frac{\langle R_{\rm g}^2\rangle - \langle R_{\rm g}\rangle^2}{\langle R_{\rm g}^2\rangle}\right)^{\frac{1}{2}} \, .
\label{eq:fluc}
\end{equation}
This is demonstrated in Fig. \ref{fgr:asph-fluc} for both SCNPs synthesized from linear chains and ring polymers. 
Since the internal mobility reflected by $\delta$ is relevant in the context of potential functionality, this correlation indicates a connection between shape and function and motivates our choice to classify degree of disorder in SCNPs in terms of shape parameters. Furthermore, we find that SCNPs synthesized from ring polymer precursors are generally less deformable -- i.e they exhibit smaller $\delta$ -- than those synthesized from linear chains. 

\begin{figure}
  \includegraphics[height=8cm]{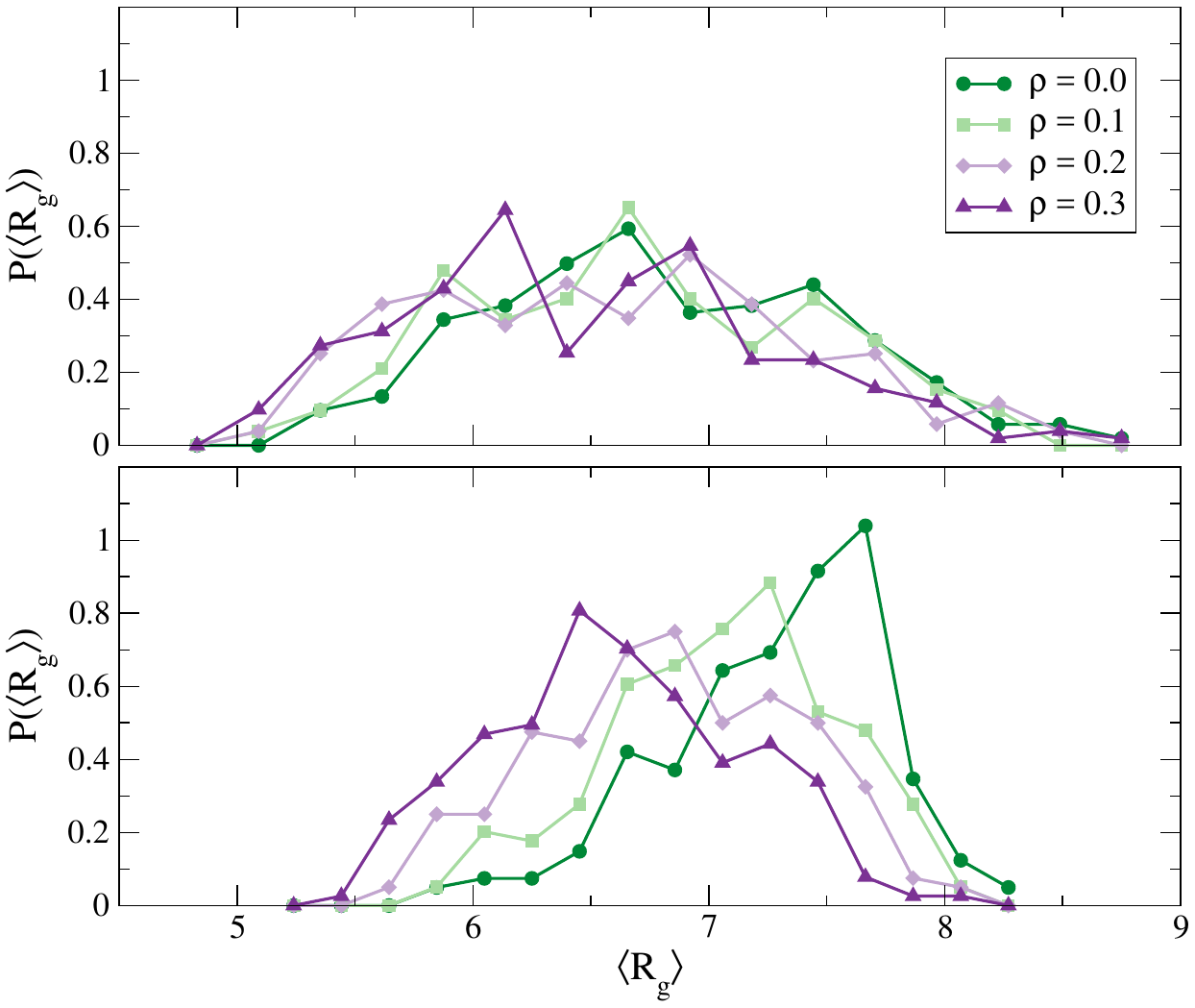}
  \caption{Distribution of radius of gyration $R_g$, at infinite dilution, for SCNPs synthesized from linear (a) and ring (b) precursors at various densitites. The inset shows the cumulative distribution function for the linear case. Brackets denote time-averages over the trajectory of a single SCNP. Symbol codes have the same meaning in both panels.}
  \label{fgr:rgyr-hist}
\end{figure}

Fig. \ref{fgr:rgyr-hist} shows the distribution of the time-averaged radius of gyration $\langle R_g \rangle$ at infinite dilution for SCNPs synthesized from chains and rings at various densities. In the case of the linear precursors, we cannot infer a clear trend from the distribution of SCNP size upon increasing the density at which the synthesis is performed. Still, we do see some drift to lower $\langle R_g \rangle$ from the cumulative distribution (Fig. \ref{fgr:rgyr-hist}a, inset). On the other hand,  in the case of the ring polymer precursors, we see a clear shift in the maximum of the distribution of SCNP size towards lower $\langle R_g \rangle$ along with a reduction of its asymmetry. 

\begin{figure}
\centering
  \includegraphics[height=7.5cm]{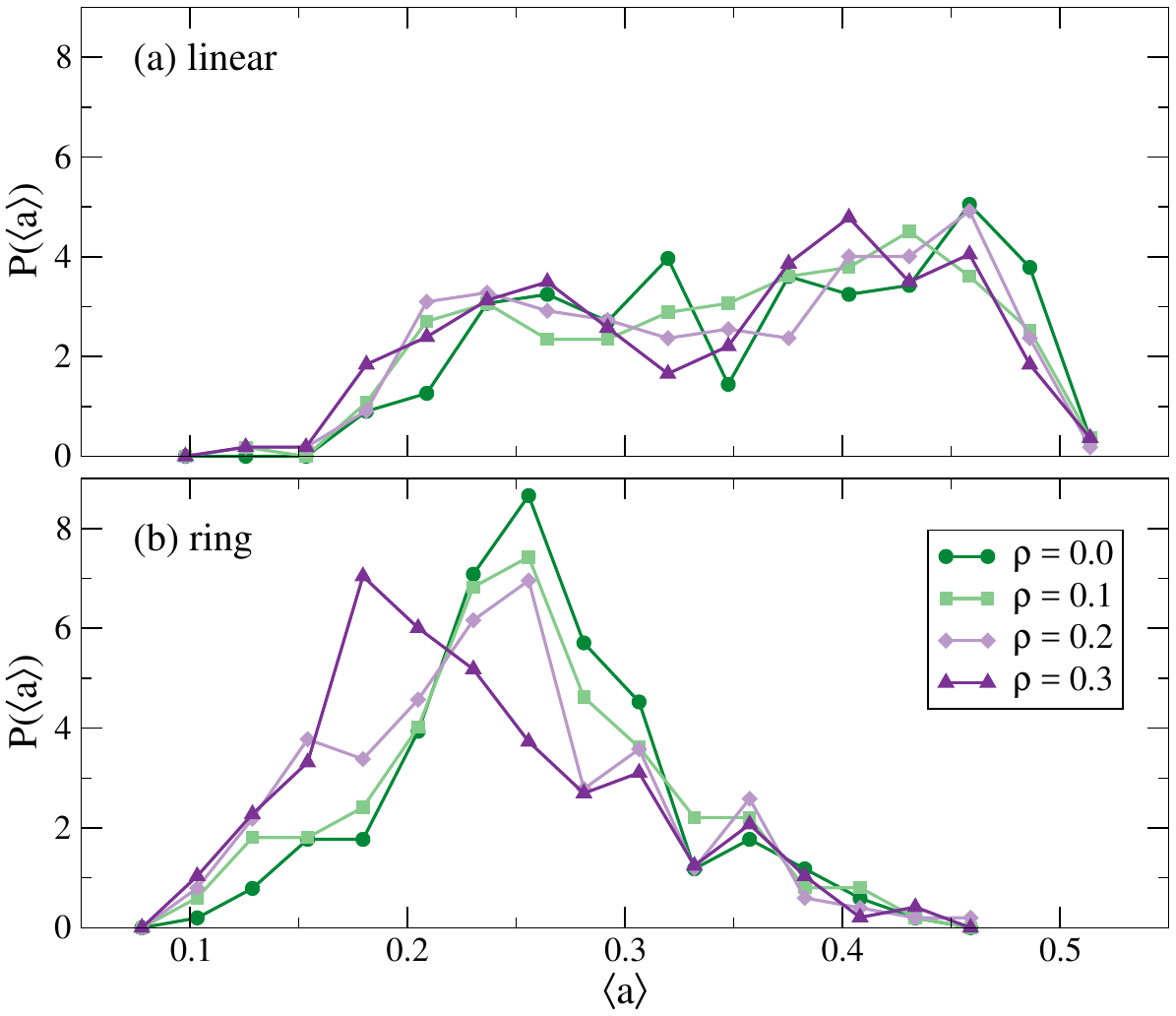}
  \caption{Distribution of asphericity $a$, at infinite dilution, for SCNPs synthesized from linear (a) and ring (b) precursors at various densitites. Brackets denote time-averages over the trajectory of a single SCNP. Symbol codes have the same meaning in both panels.}
  \label{fgr:asph-hist}
\end{figure}

\begin{figure}
\centering
  \includegraphics[height=5.5cm]{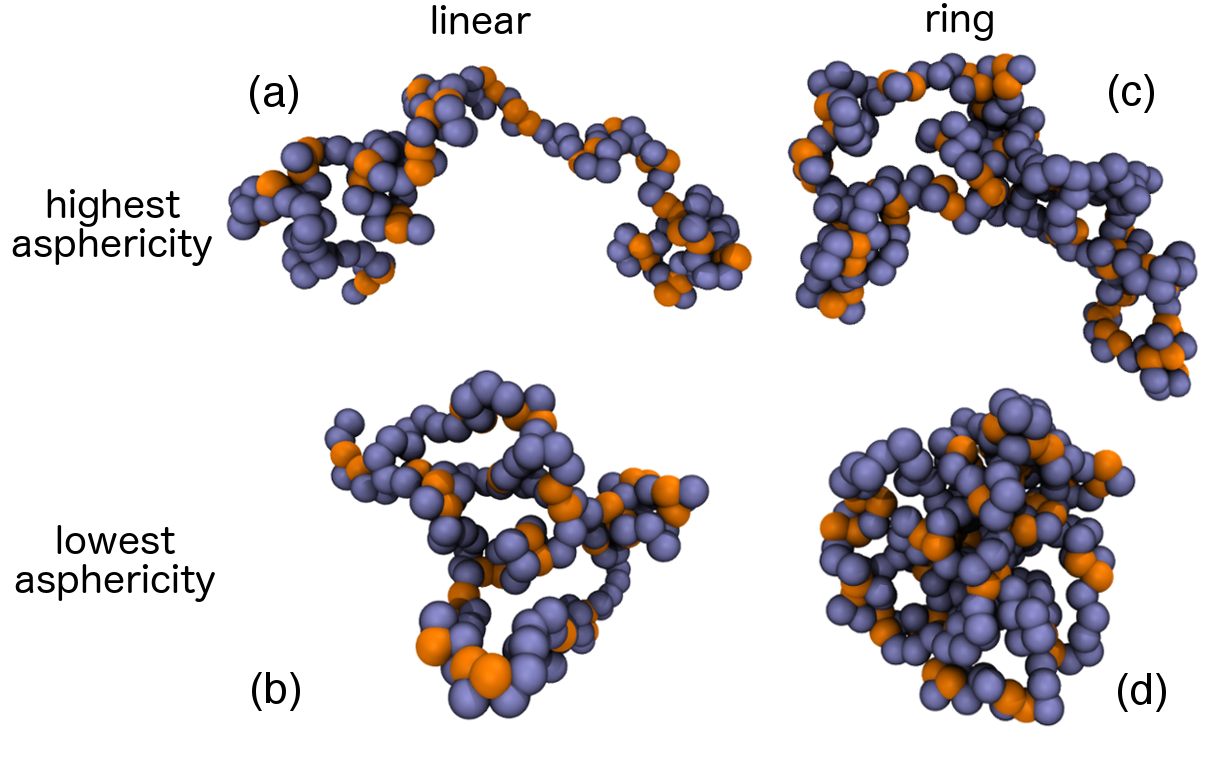}
  \caption{Representative snapshots, at infinite dilution, of SCNPs synthesized from linear (a, b) and ring precursors (c, d) at a density of $\rho = 0.3$. The selected SCNPs belong to the 10\% with the highest (a, c) and the 10\% with the lowest (b, d) asphericity. Reactive monomers are colored in orange.}
  \label{fgr:snapshots}
\end{figure}

The former observations go along with a change of the shape of the resulting ring SCNPs towards more spherical conformations, as can be seen from the distribution of asphericity $\langle a \rangle$ in Fig.~\ref{fgr:asph-hist}b. No significant changes in asphericity are found in the case of the linear SCNPs (Fig.~\ref{fgr:asph-hist}a). It should be noted that both shape parameters, $\langle R_g \rangle$ and $\langle a \rangle$, exhibit a very broad distribution across all densities, demonstrating the intrinsic structural and topological polydispersity of SCNPs, which appears to be preserved when carrying out synthesis under crowding conditions. To illustrate this, we include snapshots of representative conformations of SCNPs of both high and low asphericities (Fig.~\ref{fgr:snapshots}).     

\begin{figure}
\centering
  \includegraphics[height=5.7cm]{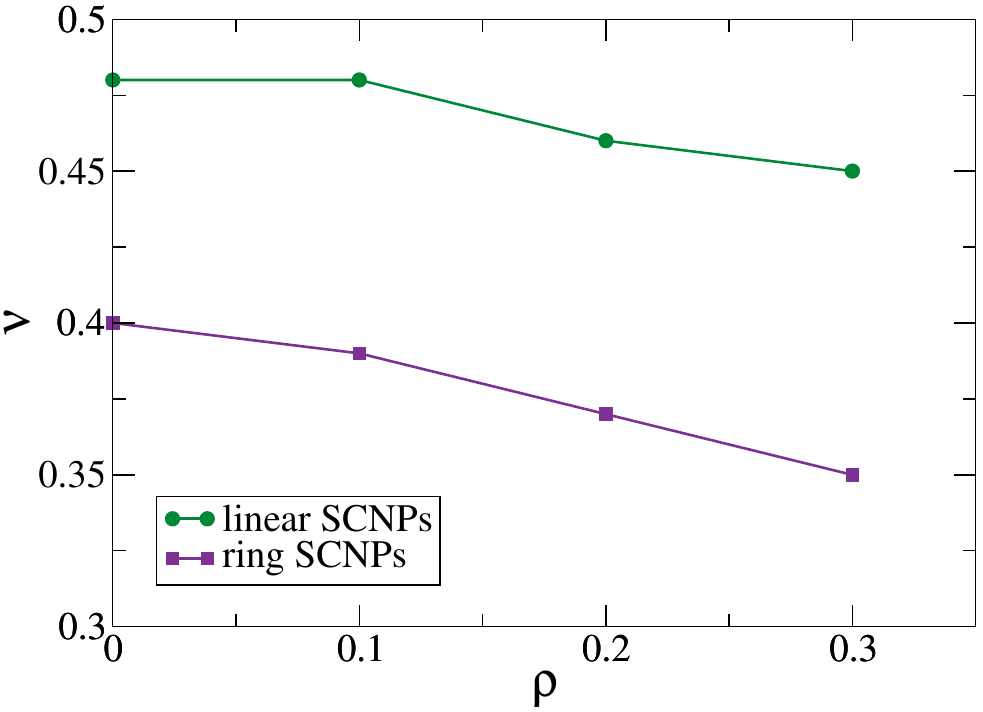}
  \caption{Scaling exponents of the whole ensemble of swollen ($\rho = 0$) SCNPs synthesized from either linear (green) or ring (purple) polymer precursors at densities ranging from $\rho = 0$ to $\rho = 0.3$.}
  \label{fgr:scaling-SCNP}
\end{figure}

Another way to gain insight on the average intramolecular structure of the SCNPs is to analyze their intramolecular form factors: 
\begin{equation}
w(q) = \left\langle \frac{1}{N} \sum_{j,k} \exp\left[ i \mathbf{q}\cdot ( \mathbf{r}_j - \mathbf{r}_k )\right] \right\rangle \, , 
\label{eq:formfactor}
\end{equation}
where $\mathbf{q}$ is the wave vector and the sum is restricted over monomers belonging to the same SCNP. In the fractal regime, $1/R_g \lesssim q \lesssim 1/b $, where $b$ is the bond length, the form factor is expected to scale as $w(q) \sim q^{-1/\nu}$, with $\nu$ the scaling exponent.\cite{Rubinstein2003}
Fig.~\ref{fgr:scaling-SCNP} displays the scaling exponents  of the SCNPs in the swollen state ($\rho = 0$) as a function of the crowding density at synthesis. We notice that the structures of SCNPs with different precursor topologies differ already when synthesized at infinite dilution, even though their precursors exhibit the same Flory-like scaling behavior at $\rho \rightarrow 0$ ($\nu \approx\nu_F = 0.59$, see Fig.~\ref{fgr:scaling-precursor}). Increasing the density of crowding molecules for the synthesis leads to a decrease in the scaling exponents for both precursor topologies, which signifies that the SCNPs synthesized under crowding conditions adopt more compact structures even in their swollen state after removing the crowders.

The scaling exponent of the SCNPs synthesized from linear precursors at $\rho \rightarrow 0$,  $\nu = 0.48$  (Fig.~\ref{fgr:scaling-SCNP}), is similar to that of linear polymers in melts or $\theta$-solvents ($\nu = 1/2$). This finding can be attributed to the self-avoiding character of the precursor chain under dilute good solvent conditions, which promotes cross-linking between monomers separated by small contour distances, resulting in globulation only at local scales. As has been previously reported\cite{Moreno2013, LoVerso2014}, however, the efficient mechanism of global compaction of SCNPs is the formation of bonds across long contour distances. In linear precursors this constitutes an unfrequent event, happening mostly towards the end of the synthesis, when some distant unlinked reactive groups are still present and the polymer backbone has to undergo large reorientations to bring them into contact. Although ring polymers exhibit the same self-avoiding behavior at high dilution as linear chains, their intrinsic topology makes cross-linking over long contour distances -- and thus global compaction -- much more likely for the same molecular size of the precursor. As a consequence, SCNPs obtained from ring precursors
at $\rho \rightarrow 0$ show a lower exponent, $\nu \approx 0.4$ than their counterparts synthesized from linear chains in the same conditions.

Upon an increase in the crowding density at synthesis we observe a small but consistent decrease in the scaling exponents of the swollen SCNPs of both precursor types. Together with the distributions of asphericity and radius of gyration, these results show that synthesis under crowding conditions leads to more compact and globular SCNPs than synthesis at high dilution. 
Interestingly, after removing the crowders ($\rho = 0$), the ring SCNPs essentially retain the scaling exponents displayed by the unlinked ring  precursors at the corresponding densities of the synthesis (see Figs.~\ref{fgr:scaling-precursor} and \ref{fgr:scaling-SCNP}). This behaviour suggests an analogy between the crumpled globular state of a ring polymer in a melt or a concentrated solution ($\nu \gtrsim 1/3$) and the cross-linked conformation of a ring SCNP in dilute conditions. The crumpled globular conformations adopted by rings in the former crowded environments are characterized by each subchain of the ring being condensed in itself\cite{Grosberg1988, Lua2004, Bohn2010}. One may argue that such conformations allow the precursor to fully undergo cross-linking without the need for large reorientations as is the case for linear precursors. Thus, the formation of permanent loops in the SCNP would freeze it in a typical conformation of its ring precursor and allow the SCNP to retain the crumpled globular conformation, after removing the crowders, in the swollen state. 


\begin{figure*}
 \centering
 \includegraphics[height=9.7cm]{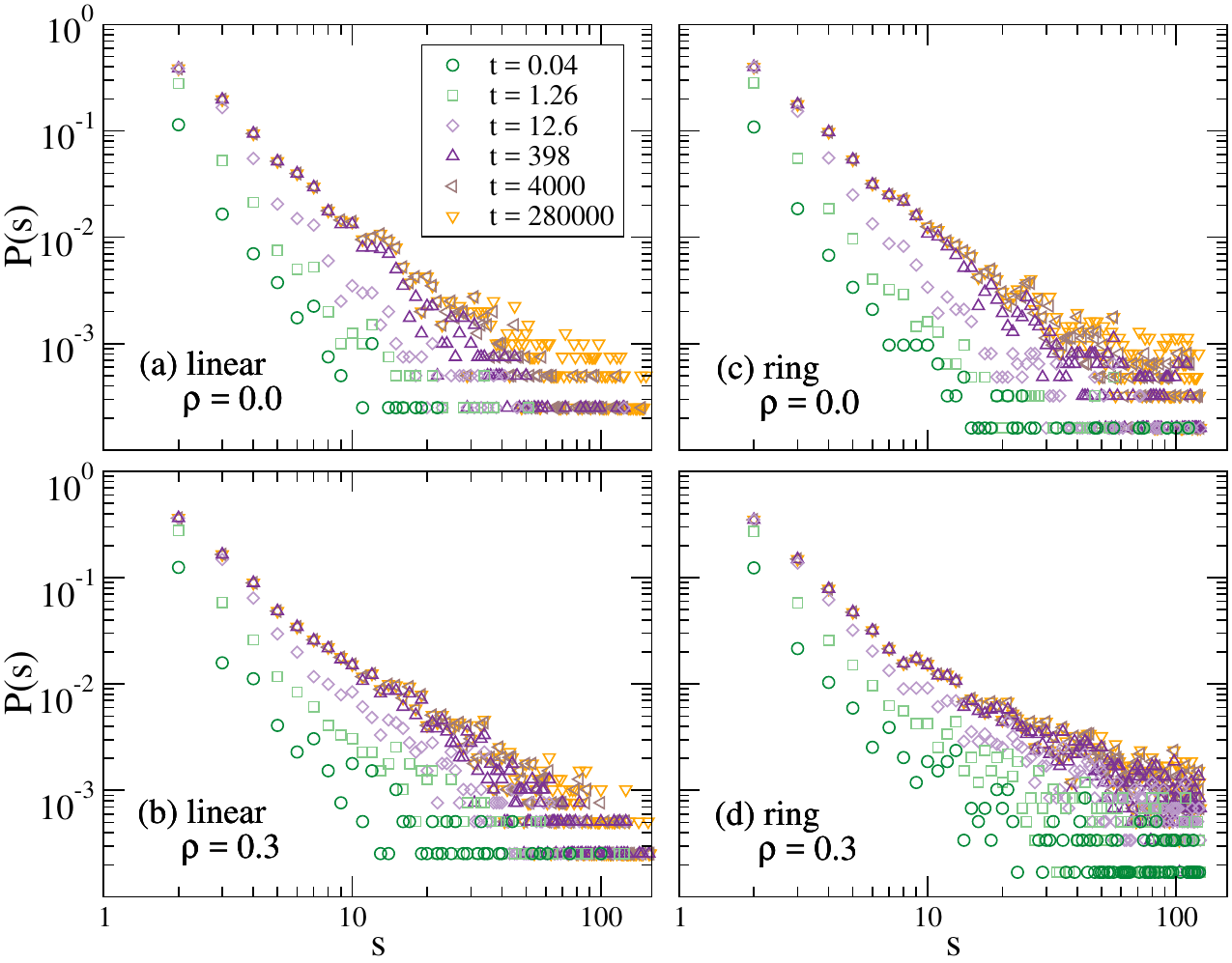}
 \caption{Time evolution of the histogram of contour distances $s$ between bonded reactive groups, for the SCNPs synthesized from linear precursors at infinite dilution (a) and $\rho = 0.3$ (b), as well as of SCNPs synthesized from ring precursors at infinite dilution (c) and $\rho = 0.3$ (d). Different data sets correspond to different selected times (see legend). At the latest time (orange), all SCNPs were fully cross-linked.}
 \label{fgr:contour-hist}
\end{figure*}

\begin{figure*}
 \centering
 \includegraphics[height=8.5cm]{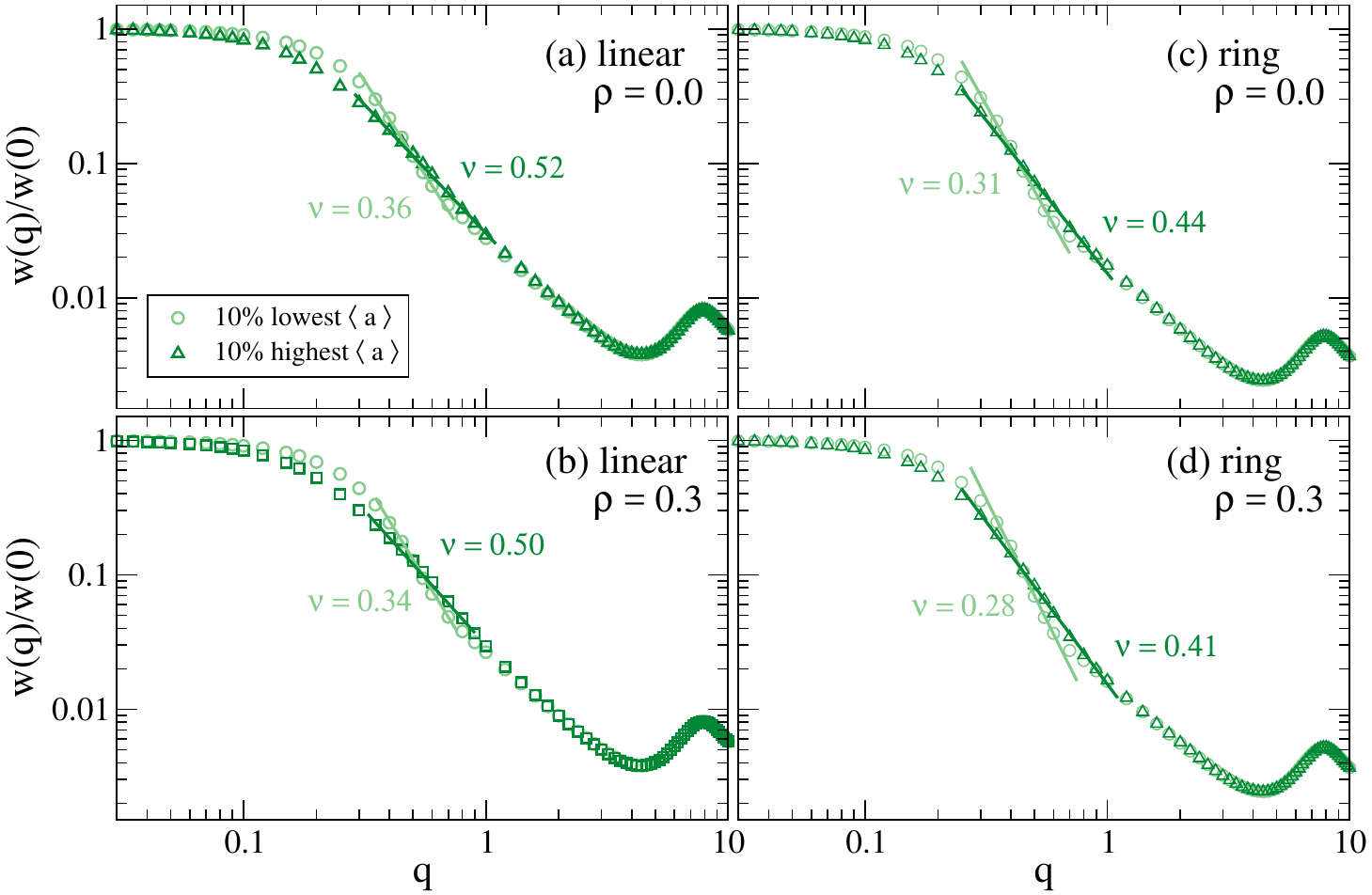}
 \caption{Normalized form factors, at infinite dilution, for the 10\% with the lowest and 10\% with the highest asphericity $a$ of SCNPs synthesized from linear precursors at infinite dilution (a) and $\rho = 0.3$ (b), as well as of SCNPs synthesized from ring precursors at infinite dilution (c) and $\rho = 0.3$ (d). Solid lines are fits to power-laws, $w(q) \sim q^{-1/\nu}$, in the fractal regime. Each fitted line is annotated with its scaling exponent $\nu$. Symbol codes have the same meaning across all panels.}
 \label{fgr:form-factors}
\end{figure*}

If this is the case for SCNP rings, one should expect to see the formation of bonds involving long contour distances already at the beginning of the cross-linking process instead of only towards the end, as was reported for SCNPs synthesized from linear precursors\cite{Moreno2013}. To confirm this assumption, we calculate the contour distances $s = |i-j|$ between bonded reactive groups
($i,j$) at different times of the cross-linking process. Fig.~\ref{fgr:contour-hist} shows the time evolution of the histogram of such contour distances, $P(s)$, from the beginning of the cross-linking until all SCNPs are fully cross-linked. We observe that $P(s)$ is a monotonically decreasing function of $s$ for linear precursors, while it exhibits a plateau at very large $s$ for ring precursors. At short contour distances, $s \lesssim 10$, the time evolution of the histogram is qualitatively the same for all cases, with  $P(s)$ growing up until $t \sim 400$, after which no significant increase is observed. At large contour distances, $s \gtrsim 40$ however, the histograms exhibit large qualitative differences. While for the linear precursors in dilute conditions (Fig. \ref{fgr:contour-hist}a), bonds begin to form significantly at $s \gtrsim 40$ only after $t \sim 400$, they are formed right from the beginning in the ring precursors at $\rho = 0.3$ and show the strongest growth in an intermediate regime $0 \lesssim t \lesssim 400$ (Fig. \ref{fgr:contour-hist}d). This finding supports our assumption that the crowding conditions allow for the freezing of the ring SCNPs, through the formation of some permanent long-range loops, into the crumpled globular conformations of their ring precursors, which are retained by the SCNP upon removing the crowders.


Because of the topological and structural polydispersity of the SCNPs, it is instructive to consider subsets of the synthesized SCNPs with similar structural features and analyze their conformations separately. To this end, we select the $10\%$ most globular and least globular SCNPs, according to their average asphericity, and calculate their intramolecular form factors separately. Fig.~\ref{fgr:form-factors} displays the form factors (at infinite dilution) of these two subsets of SCNPs synthesized from linear chains (left column) and ring polymers (right column). We show results for the synthesis at infinite dilution (top) and a density of $\rho = 0.3$ (bottom). An inspection of the scaling behavior of the form factor in the fractal regime reveals that the two subsets considered not only differ in their shape but also in their degree of compaction (as revealed by the lower $\nu$ for lower asphericity). For both SCNPs synthesized from ring and linear precursors, we can find individual molecules with aspherical structure and sparse topology, as well as others with spherical and compact one. When going from infinite dilution to $\rho = 0.3$ at synthesis, the scaling exponents of both the most and least globular SCNPs decrease to about the same extent as the average scaling exponents (Fig.~\ref{fgr:scaling-SCNP}). It is worthy of remark that the scaling exponents of the $10\%$ of ring SCNPs with the lowest asphericity lie below the value expected for globular objects ($\nu = 1/3$).
Actually, this is just a consequence of approaching the limit of Porod scattering in the form factor ($w(q) \sim q^{-4}$), 
which yields an {\it effective} exponent $\nu = 1/4$. \cite{Rubinstein2003,Grest1987, Grest1989} 
This limit is almost reached for the $10\%$ most spherical SCNPs synthesized from rings at $\rho = 0.3$  (Fig.~\ref{fgr:form-factors}d).
The observation of Porod scattering is a manifestation of the highly spherical and dense, unpenetrable character of this subset of SCNPs.

\begin{figure}
\centering
  \includegraphics[height=7.5cm]{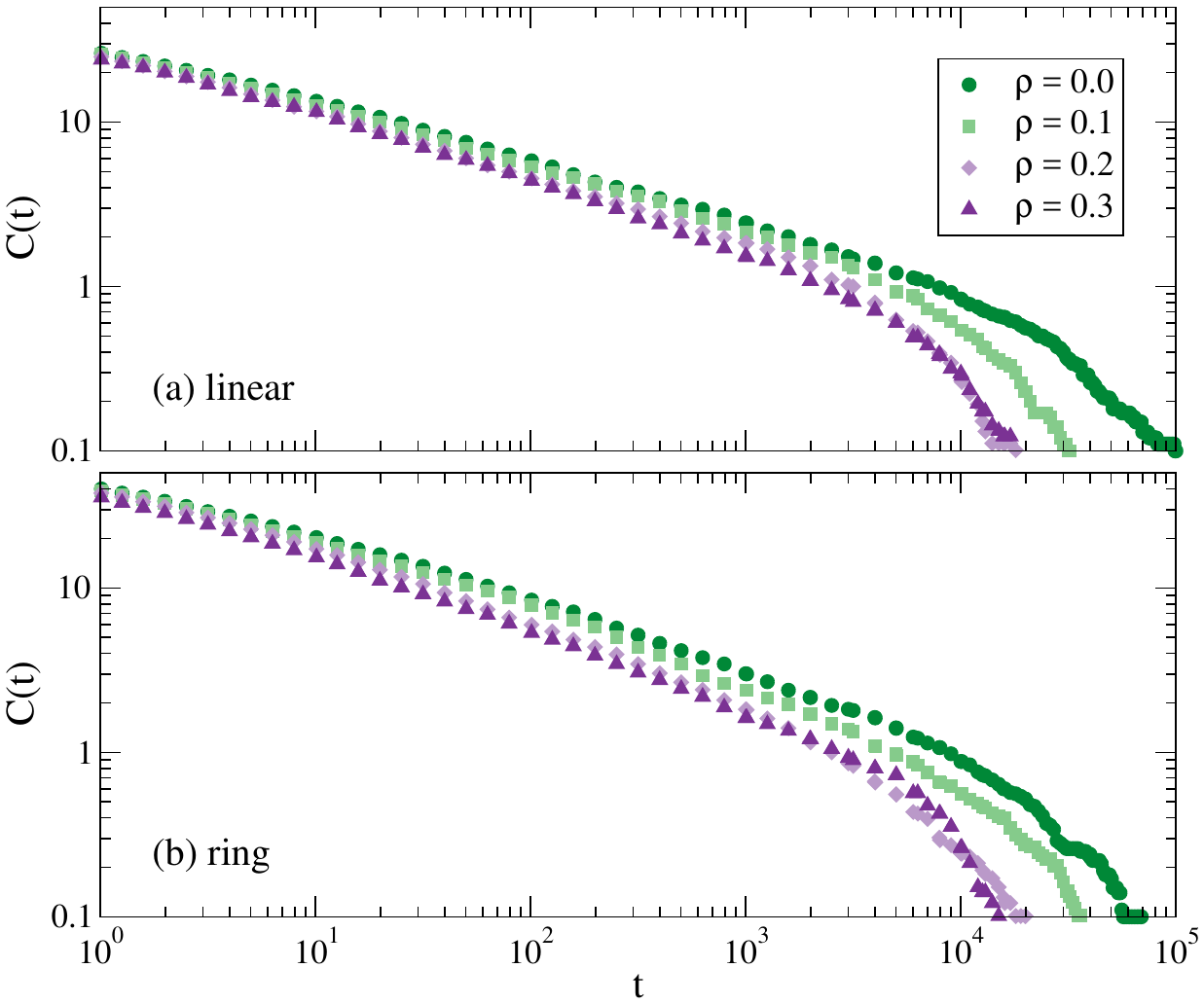}
  \caption{Number of unlinked reactive monomers per reactive molecule $C(t)$ versus time $t$ for SCNPs synthesized from linear (a) and ring (b) precursors at various densitites. Symbol codes have the same meaning in both panels.}
  \label{fgr:crosslink-rate}
\end{figure}

To conclude this section, we discuss the effect of crowding on the cross-linking rate for SCNP synthesis. Fig.~\ref{fgr:crosslink-rate} shows the ensemble-average number of unlinked reactive monomers $C(t)$ as a function of time. Although the total number of reactive monomers is higher for the ring precursors ($N_r = 62$ versus $N_r = 40$ for linear precursors), their cross-linking process is faster. Furthermore, increasing the density of non-reactive crowders accelerates cross-linking up to the highest density considered. However, the overlap of the curves for the two highest densities, $\rho = 0.2$ and $\rho=0.3$, suggests that there shoud be a final  crossover to the opposite effect (deceleration) when going to even higher densities. In the case of the linear precursors, the crossover should be expected when going beyond the entanglement density \cite{Rubinstein2003} 
$\rho_{\rm e} = (N_{\rm e}/N)^{3\nu_{\rm F}-1} \approx 0.5$, where $N_{\rm e} \approx 65$ is the entanglement length \cite{Ever_pp_procedure}. Once entanglements are present in the system, the lateral confinement of the individual chains, which become forced to reptate along their primitive paths, should slow down the cross-linking process. An exploration of densities beyond the entanglement concentration is beyond the scope of this work. Still it must be noted that, in addition to the growing computational cost of simulating higher concentrations, the relaxation time of linear polymers on length scales larger than the tube diameter will show a steep increase. This relaxation time is provided by the reptation model and scales as \cite{Rubinstein2003} $\tau_{\rm rep} \sim \rho^{3(1-\nu_{\rm F})/(3\nu_{\rm F}-1)}N^3 \sim \rho^{1.6}N^3$. This is a much stronger dependence than the Rouse scaling $\sim N^2$ for untentangled chains, and as aforementioned, would  dramatically increase the duration of the cross-linking process. Moreoever, concatenations during the cross-linking of the ring precursors with their respective crowders, as illustrated in Fig.~\ref{fgr:concatenation}, will presumably increase at higher densities because of increasing interpenetration
of the polymers.

\section{Conclusions}
In summary, by means of molecular dynamics simulations we have investigated the effects of the precursor topology, as well as carrying out the synthesis under crowding conditions, on the structural and topological properties of single-chain nanoparticles. To this end, the cross-linking process of either ring or linear precursors with randomly distributed reactive groups has been simulated in the presence of non-reactive molecules of the same topology, to study the effects of purely steric, non-specific interactions in a range of densities, $\rho = 0.1 - 0.3$, that is typical of, e.g., cellular environments\cite{Theillet2014}.   
We have shown that using ring polymers as precursor molecules constitutes a promising new route for the design of compact and globular SCNPs, an objective that has remained challenging with the standard synthesis protocols, which have been shown to rather result in open sparse objects\cite{Pomposo2014a, Moreno2013, LoVerso2014}. 

Furthermore, increasing the density of crowder molecules present at the time of synthesis leads to a compaction of the resulting SCNPs, which is accompanied by a shift towards more spherical conformations when using ring polymer precursors. The scaling exponents found in the swollen state (high dilution) for ring SCNPs are essentially identical to the exponents of the ring precursors at the density of synthesis. Thus, the swollen SCNPs retain, through permanent long-range cross-links, the crumpled globular conformations of the crowded rings. We conclude that the intrinsic topology and the different collapse behaviour of linear polymers and ring polymers in crowded solutions explain why ring precursors lead to more globular and compact SCNPs. 

Our findings are relevant for the design of possible new synthesis routes involving different precursor topologies or crowded solutions. So far most of the protocols for the synthesis of SCNPs have been limited to linear precursors containing bulky side groups or branches\cite{Harth2002, Jiang2005, Croce2007, Luzuriaga2010, Sanchez-Sanchez2013a}. Our simulations suggest that ring polymers as precursor molecules are promising canditates for the synthesis of globular soft nanoparticles for applications in nanomedicine and catalysis. While the synthesis and purification of monodisperse, unknotted and nonconcatenated rings without linear contaminants remains challenging, recent advances in isolation of ring polymers from linear chains of the same molecular weight\cite{Pasch1999, Lee2000, Lee2002, Tezuka2002, Montenegro-Burke2016, Barroso-Bujans2017} suggest that the synthesis of SCNPs from ring polymer precursors will be experimentally realizable in the near future.  

\section{Acknowledgements}

We acknowledge financial support from the projects MAT2015-63704-P (MINECO-Spain and FEDER-UE) and IT-654-13 (Basque Government, Spain).


\providecommand*{\mcitethebibliography}{\thebibliography}
\csname @ifundefined\endcsname{endmcitethebibliography}
{\let\endmcitethebibliography\endthebibliography}{}

\end{document}